\documentclass[aps,prl,twocolumn,floatfix, superscriptaddress,nofootinbib]{revtex4-2} 

\usepackage{amsmath,amssymb}
\usepackage{braket}
\usepackage{mathtools}
\usepackage{xcolor, ulem}
\usepackage{siunitx}
\usepackage{CJK} 
\usepackage{lipsum}
\usepackage{graphicx}
\usepackage{footnotebackref}

\definecolor{pal0}{rgb}{0.8941, 0.102 , 0.1098}
\definecolor{pal1}{rgb}{0.2157, 0.4941, 0.7216}
\definecolor{pal2}{rgb}{0.302 , 0.6863, 0.2902}
\definecolor{pal3}{rgb}{0.5961, 0.3059, 0.6392}
\definecolor{pal4}{rgb}{1.    , 0.498 , 0.    }

\usepackage{algorithm}
\usepackage[noend]{algpseudocode}
\usepackage{setspace}

\usepackage{dsfont}

\newcommand{\R}{\mathbb{R}}
\newcommand{\N}{\mathbb{N}}

\renewcommand{\v}[1]{\boldsymbol{#1}}

\def \k{{\mathbf k}}
\def \q{{\mathbf q}}
\def \g{{\mathbf g}}
\def \R{{\mathbf R}}
\def \r{{\mathbf r}}
\def \K{{\mathbf K}}
\def \G{{\mathbf G}}
\def \A{\mathbf{A}}

\newcommand{\m}[1]{\mathsf{#1}}

\usepackage{pifont}

\begin{document}

    \begin{CJK*}{UTF8}{min}

	\title{Strain-induced quantum phase transitions in magic angle graphene}
	
\author{Daniel E. Parker}
\affiliation{Department of Physics, Harvard University, Cambridge, MA 02138, USA}
\affiliation{Department of Physics, University of California, Berkeley, CA 94720, USA}
 \author{Tomohiro Soejima (副島智大)}
\affiliation{Department of Physics, University of California, Berkeley, CA 94720, USA}
\author{Johannes Hauschild}
\affiliation{Department of Physics, University of California, Berkeley, CA 94720, USA}
\author{Michael P. Zaletel}
\affiliation{Department of Physics, University of California, Berkeley, CA 94720, USA}
\affiliation{Materials Sciences Division, Lawrence Berkeley National Laboratory, Berkeley, California 94720, USA}
\author{Nick Bultinck}
\affiliation{Department of Physics, University of California, Berkeley, CA 94720, USA}
\affiliation{Department of Physics, Ghent university, 9000 Ghent, Belgium}
	
		\date{\today}
	
	\begin{abstract}
We investigate the effect of uniaxial heterostrain on the interacting phase diagram of magic-angle twisted bilayer graphene.
Using both self-consistent Hartree-Fock and density-matrix renormalization group calculations, we find that  small strain values ($\epsilon \sim{} 0.1-0.2\%$) drive a zero-temperature phase transition between the symmetry-broken ``Kramers intervalley-coherent'' insulator and a nematic semi-metal.
The critical strain lies within the range of experimentally observed strain values, and we therefore predict that strain is at least partly responsible for the sample-dependent experimental observations.

	\end{abstract}
	\maketitle
	
	\end{CJK*}

Experiments on different twisted bilayer graphene (TBG) devices, all close to the first magic angle, have produced a broad variety of different low-temperature phase diagrams. For example, at the charge neutrality point (CNP), both semi-metallic \cite{PabloMott,PabloSC,Dean-Young,PabloNematic,LiVafekscreening,HerreroFlavourHund} and insulating \cite{sharpe2019emergent,YoungQAH,efetov,EfetovScreening,AndreiChern} states have been observed. The insulating devices are thought to be divided into two groups. In the first group \cite{sharpe2019emergent,YoungQAH}, one of the graphene sheets is almost perfectly aligned with the hexagonal Boron-Nitride (hBN) substrate, which breaks the two-fold rotation symmetry and therefore generates mass terms for the Dirac cones \cite{Hunt13,Amet13,Zibrov,Jung,YankowitzJung,Kim2018} in the single-particle continuum model of TBG \cite{MacDonald2011,Morell2010,Santos}. In the second group of devices \cite{efetov,AndreiChern}, those without substrate alignment, the Coulomb interaction is believed to be responsible for the insulating behavior. Both analytical and numerical studies \cite{XieMacDonald, KIVCpaper} of pristine TBG at the CNP indeed find an insulating ground state, due to spontaneous ``Kramers inter-valley coherent'' (KIVC) order \cite{KIVCpaper}. The KIVC state is thus a promising candidate for the CNP insulators in Ref. \cite{efetov}, as well as the $|\nu| = 2$ insulators in general, but cannot explain the semimetals observed in Refs. \cite{PabloMott,Dean-Young,PabloNematic,LiVafekscreening,HerreroFlavourHund}. Moreover, self-consistent Hartree-Fock (SCHF) predicts a KIVC gap of $\sim \SI{20}{\milli\electronvolt}$ \cite{KIVCpaper}, while experiments measure a global transport gap of only $\sim \SI{1}{\milli\electronvolt}$ \cite{efetov}. 

An important question is thus: what weakens the insulators in some experimental devices, and destroys them in others? Twist-angle disorder is expected to be at least partly responsible for this \cite{Zeldov,CascadeShahal,WilsonPixley,NeupertRyu}. Another possible culprit is the presence of strain in the graphene sheets. Uniaxial heterostrain is characterized by a parameter $\epsilon$, which scanning tunneling spectroscopy experiments have found to be in the range $\epsilon = 0.1 - 0.7 \%$ \cite{ColumbiaSTM,CaltechSTM,PrincetonSTM}. Although these values seem small at face value, strain contributes to the Hamiltonian as a perturbation of order $\epsilon \hbar v_F/a$, which is $\sim 20$ meV for $\epsilon = 0.5\%$ --- precisely the energy scale at issue. Further evidence for the importance of strain comes from symmetry considerations. In the absence of strain, models at even integer filling show that although the ground state has KIVC order, there is a close competitor whose energy is only slightly higher: a nematic semi-metal \cite{CaltechSTM,liu2020Nematic,KIVCpaper,KangVafek,DMRGpaper}. As elucidated in Ref.~\cite{liu2020Nematic}, the semi-metal has two Dirac points close to, but not at, the mini-BZ $\Gamma$ point, spontaneously breaking the three-fold rotational symmetry $C_{3z}$. The shear part of uniaxial strain breaks the $C_{3z}$ symmetry, and thus one expects on general grounds that strain will lower the energy of the nematic semi-metal relative to the rotationally invariant insulating states.
However, despite this expectation, Refs. \cite{liu2020Nematic,KIVCpaper} found that if strain is modeled using the phenomenological method of Ref.~\cite{PoSenthilC3}, it cannot stabilize the semi-metal. 

This work provides a careful treatment of the effects of strain on the correlated insulators using a more realistic model for strained TBG \cite{BiFu}. We find that physical strain values can drive a zero-temperature phase transition from the KIVC insulator to a semi-metal at even integer fillings. Our results at charge neutrality are obtained using SCHF, and our results at $\nu = -2$ ($\nu$ is the number of electrons per moir\'e unit cell relative to charge neutrality) using both density-matrix renormalization group (DMRG) and SCHF. Our application of DMRG to TBG is a technical advance in its own right, as it is the first DMRG study to keep both valley degrees of freedom, which is essential for correctly identifying the even-integer insulators. Similar to earlier works on single-valley models \cite{KangVafek,DMRGpaper}, we find that DMRG and SCHF agree remarkably well. In particular, DMRG confirms the presence of KIVC order at $\nu = - 2$ in the absence of strain.

\textit{Continuum model with strain --} To add uniaxial strain to the Bistritzer-MacDonald (BM) continuum Hamiltonian \cite{MacDonald2011,Morell2010,Santos}, we follow Ref. \cite{BiFu}. Uniaxial strain is characterized by the following symmetric matrix:
\begin{equation}\label{S}
    \m{S} = \left(\begin{matrix}\epsilon_{xx} & \epsilon_{xy} \\ \epsilon_{xy} & \epsilon_{yy} \end{matrix}\right) = \m{R}(\varphi)^T\left(\begin{matrix}\epsilon & \\ & -\nu_P \epsilon \end{matrix} \right)\m{R}(\varphi)\, ,
\end{equation}
where $\nu_P \approx 0.16$ is the Poisson ratio of graphene. The angle $\varphi$ corresponds to the uniaxial strain direction, and $\m{R}(\varphi)$ is a $2\times 2$ rotation matrix. Throughout this work we take $\varphi = 0$, but we have verified that our conclusions do not depend on the choice of $\varphi$. The strain magnitude is determined by the dimensionless parameter $\epsilon$, which in the devices prepared for STM study has values in the range $\epsilon = 0.1 - 0.7 \%$ \cite{ColumbiaSTM,CaltechSTM,PrincetonSTM,NadjPergeChern}. Under the combined effect of rotation and strain, the coordinates of the carbon atoms in the two graphene layers $\ell = \pm$ of TBG transform as $\R_{\ell,i}\rightarrow \left[\m{R}(\ell\theta/2)-\frac{\ell}{2}\m{S} \right]\R_{\ell,i} =: \m{M}_\ell^T \R_{\ell,i}$
where $\theta$ is the twist angle. The coordinate transformation matrix $\m{M}_\ell^T$ is correct to first order in both $\theta$ and $\epsilon$. Note that we only consider heterostrain, as it affects the electronic structure much more strongly than homostrain \cite{Heterostrain}. 

The continuum Hamiltonian in the presence of uniaxial heterostrain for the $\tau=+$ valley is given by 
\begin{equation}\label{SVham}
    H_{\tau+} = \left(\begin{matrix} D_+ & T(\r) \\ T(\r)^\dagger & D_- \end{matrix}\right)\, ,
\end{equation}
with $D_\ell$ the monolayer Dirac Hamiltonians, and $T(\r)$ the inter-layer tunneling ($H_{\tau-}$ is then fully specified by time-reversal). The Dirac Hamiltonians are given by
\begin{equation}\label{eq:strain_Dirac_Ham}
    D_\ell = -\hbar v_F\left[\m{M}_\ell(-i\boldsymbol{\nabla}+\A_\ell)-\K \right]\cdot\boldsymbol{\sigma}\, ,
\end{equation}
where $\boldsymbol{\sigma}=(\sigma_x,\sigma_y)$ are Pauli matrices acting in sublattice space, and $\K = (4\pi/3a,0)$, with $a$ the graphene lattice constant, corresponds to location of the $\tau=+$ valley. Strain shifts the locations of the Dirac points via a `vector potential' $\A_\ell = -\frac{\ell}{2}\frac{\beta\sqrt{3}}{2a}\left(\epsilon_{xx}-\epsilon_{yy}, -2\epsilon_{xy}\right)$ \cite{SuzuuraAndo,SasakiSaito}, where $\beta \sim 3.14$ characterizes the dependence of the tight-binding hopping strength on the bond length.

The inter-layer tunneling term $T(\r)$ in Eq. \eqref{SVham} has the same form as in the original BM model, albeit with differing intra and inter-sublattice interlayer tunneling amplitudes $w_{AA} = \SI{83}{\milli\electronvolt}$ and $ w_{AB} = \SI{110}{\milli\electronvolt}$ \cite{NamKoshino,KoshinoFu,CarrKaxiras}. 
To account for non-zero strain $\epsilon$, the moir\'e reciprocal lattice vectors are deformed to $\g_j = \left[\m{M}_+^{-1} - \m{M}_-^{-1}\right]\G_j$, where $\G_j$ are the reciprocal vectors of undeformed graphene.

As was shown in Ref. \cite{BiFu,Heterostrain}, uniaxial heterostrain has three important effects on the BM band spectrum: (i) while strain preserves $C_2 \mathcal{T}$ symmetry, and hence the stability of the two mini Dirac points, the three-fold rotation symmetry is broken and the two Dirac points move away from the $K^\pm$-points  towards the $\Gamma$-point in the mBZ, (ii) the two Dirac points are no longer degenerate, but are separated in energy by a few meV (thus creating small electron and hole pockets at the CNP), and (iii) the bandwidth of the `narrow' bands increases significantly -- for $\epsilon$ as small as $0.6 \%$, the bandwidth of the narrow bands is $\sim 50$ meV. Below, we investigate the effect of strain on the interacting phase diagram of TBG.

\textit{Hartree-Fock at neutrality --} 
We model interacting TBG as the BM Hamiltonian plus Coulomb interactions:
\begin{equation}\label{eq:IBM}
    H = \sum_{\k}f^\dagger_\k h(\k) f_\k + \frac{1}{2A}\sum_{\q}V_\q :\rho_\q \rho_{-\q}:\, ,
\end{equation}
where $A$ is the area of the sample, and $f^\dagger_{\k,s,\tau,m}$ creates an electron with momentum $\k$ and spin $s$ in the BM band $m$ in valley $\tau$. The charge density operators are given by $\rho_\q = \sum_\k f^\dagger_{\k}\Lambda_\q(\k)f_{\k+\q}$, where the form factor matrices $\left[\Lambda_\q(\k)\right]_{(\tau,m),(\tau',n)} = \delta_{\tau,\tau'} \langle u_{\tau,m,\k}|u_{\tau,n,\k+\q}\rangle$ are defined in terms of overlaps between the periodic part of the Bloch states of the BM Hamiltonian. The interaction is given by a gate screened Coulomb potential $V_\q = \int\mathrm{d}\r\, e^{i\q\cdot\r}V(\r) = \tanh(d_sq)[2\varepsilon_0\varepsilon_r q]^{-1}$. We work with a gate distance of $d_s = \SI{25}{\nano\meter}$, and we let the dielectric constant $\varepsilon_r$ vary between $6$ and $12$. In Eq. \eqref{eq:IBM} we also project into a subspace where most or all of the remote BM valence (conduction) bands are completely filled (empty), and $m,n$ run over only those bands whose filling is not fixed. The single-particle Hamiltonian $h(\k)$ contains the BM band energies, a HF contribution from the remote filled bands, and a subtraction term \cite{XieMacDonald,liu2020Nematic}. For more details on the definition of $h(\k)$, see Ref. \cite{DMRGpaper}.

\begin{figure}
    \centering
    \includegraphics[scale=0.24]{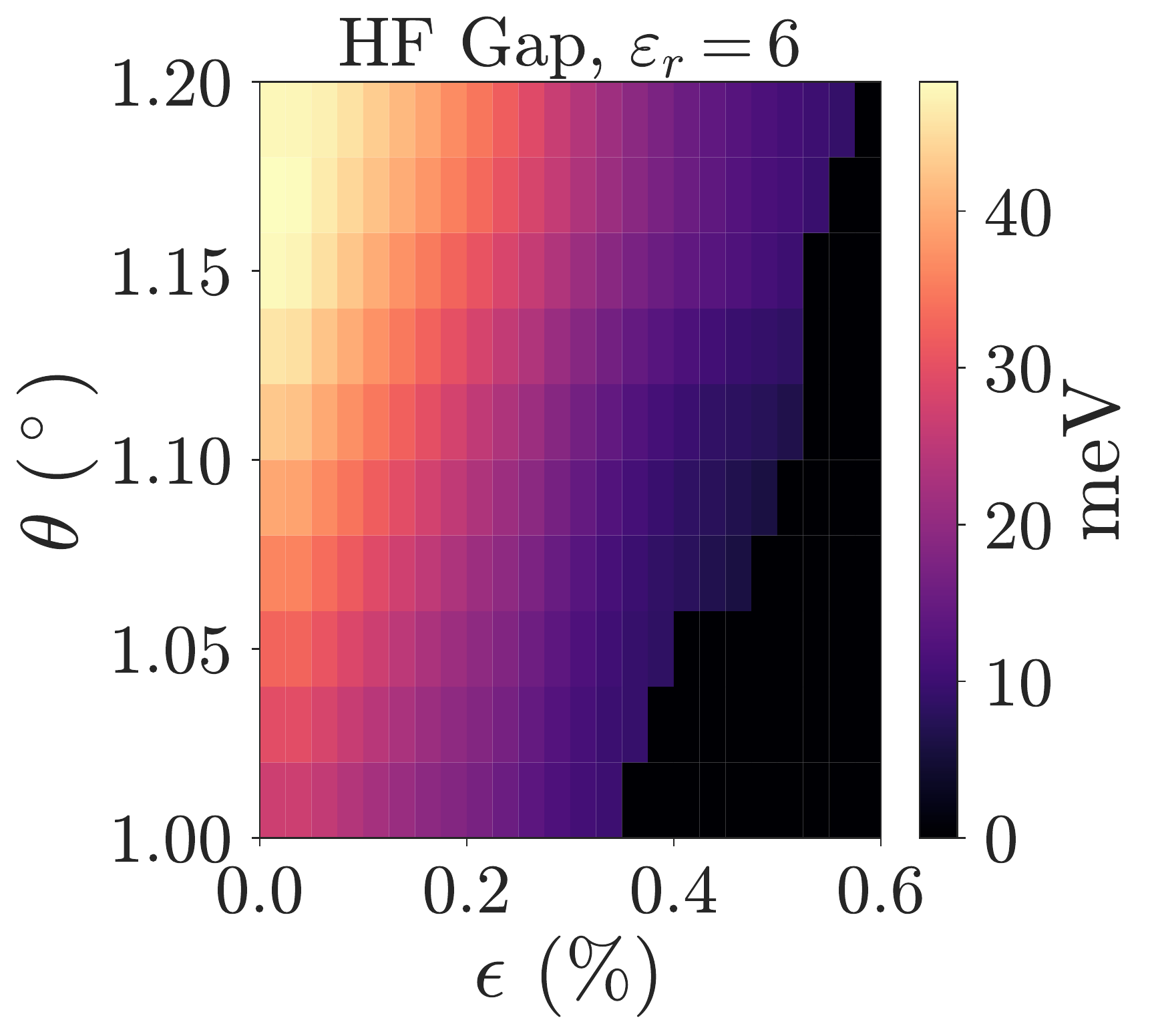}
    \includegraphics[scale=0.24]{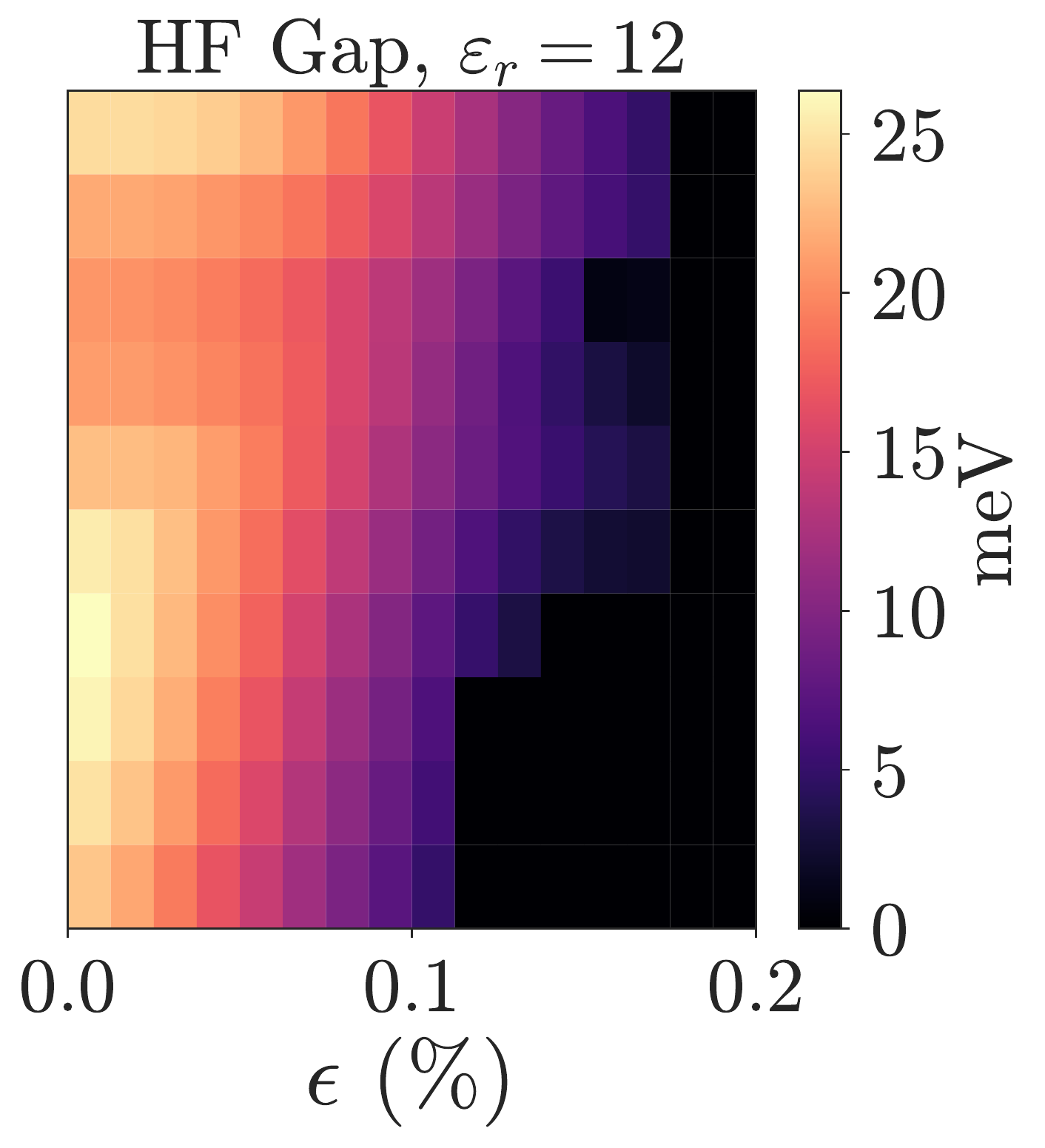}
    \caption{Particle-hole gap in the SCHF band spectrum at the CNP as a function of both twist angle $\theta$ and strain $\epsilon$, for $\varepsilon_r = 6$ (left) and $\varepsilon_r = 12$ (right). The results were obtained on a $18 \times 18$ momentum grid, keeping six bands per spin and valley. The gapped regions have KIVC order, the gapless regions correspond to a symmetric SM.
    }
    \label{fig:HFphasediag}
\end{figure}

Without strain, Ref. \cite{KIVCpaper} found that the ground state of $H$ at $\nu=-2, 0, 2$ has a charge gap and spontaneously breaks both the valley charge symmetry $e^{i\alpha\tau_z}$, and the time-reversal symmetry $\mathcal{T}= \tau_x K$, where $K$ denotes complex conjugation. However, the product $\mathcal{T}'= e^{i\pi \tau_z/2}\mathcal{T}$ is preserved. Because $\mathcal{T}' = \tau_y K$ is a (spinless) Kramers time-reversal, the insulating ground state was dubbed the Kramers inter-valley coherent (KIVC) state \cite{KIVCpaper}. 

Fig.~\ref{fig:HFphasediag} shows the HF phase diagram at the CNP as a function of twist angle and strain magnitude, for both $\varepsilon_r=6$ and $\varepsilon_r=12$. Two phases are clearly visible. The region in Fig.~\ref{fig:HFphasediag} with non-zero charge gap has KIVC order. The gapless region, on the other hand, corresponds to a semi-metal (SM) without spontaneous symmetry breaking. The HF band structure of the SM has two Dirac cones close to the $\Gamma$-point, and is therefore similar to the band structure of the strained BM Hamiltonian (for more details, see \cite{supplement}). The transition from the KIVC state to the SM in Fig. \ref{fig:HFphasediag} occurs at strain values $\epsilon \sim 0.4 - 0.6\%$ with $\varepsilon_r = 6$, and at $\epsilon \sim 0.1 - 0.2\%$ with $\varepsilon_r = 12$. These critical values lie exactly in the range of strain values observed in STM devices \cite{ColumbiaSTM,CaltechSTM,PrincetonSTM,NadjPergeChern}, from which we conclude that strain plays an important role in TBG. From Fig. \ref{fig:HFphasediag}, we also see that the KIVC state is more robust at larger $\theta$. Because at $\epsilon  = 0$ the energy difference between the KIVC state and the SM depends only weakly on $\theta$ \cite{KIVCpaper}, we attribute this feature to the fact that the active bands are less affected by strain at larger $\theta$ (in particular, the Dirac points remain further away from $\Gamma$, and the change in bandwidth is smaller).

\begin{figure}
    \centering
    \includegraphics[scale=0.35]{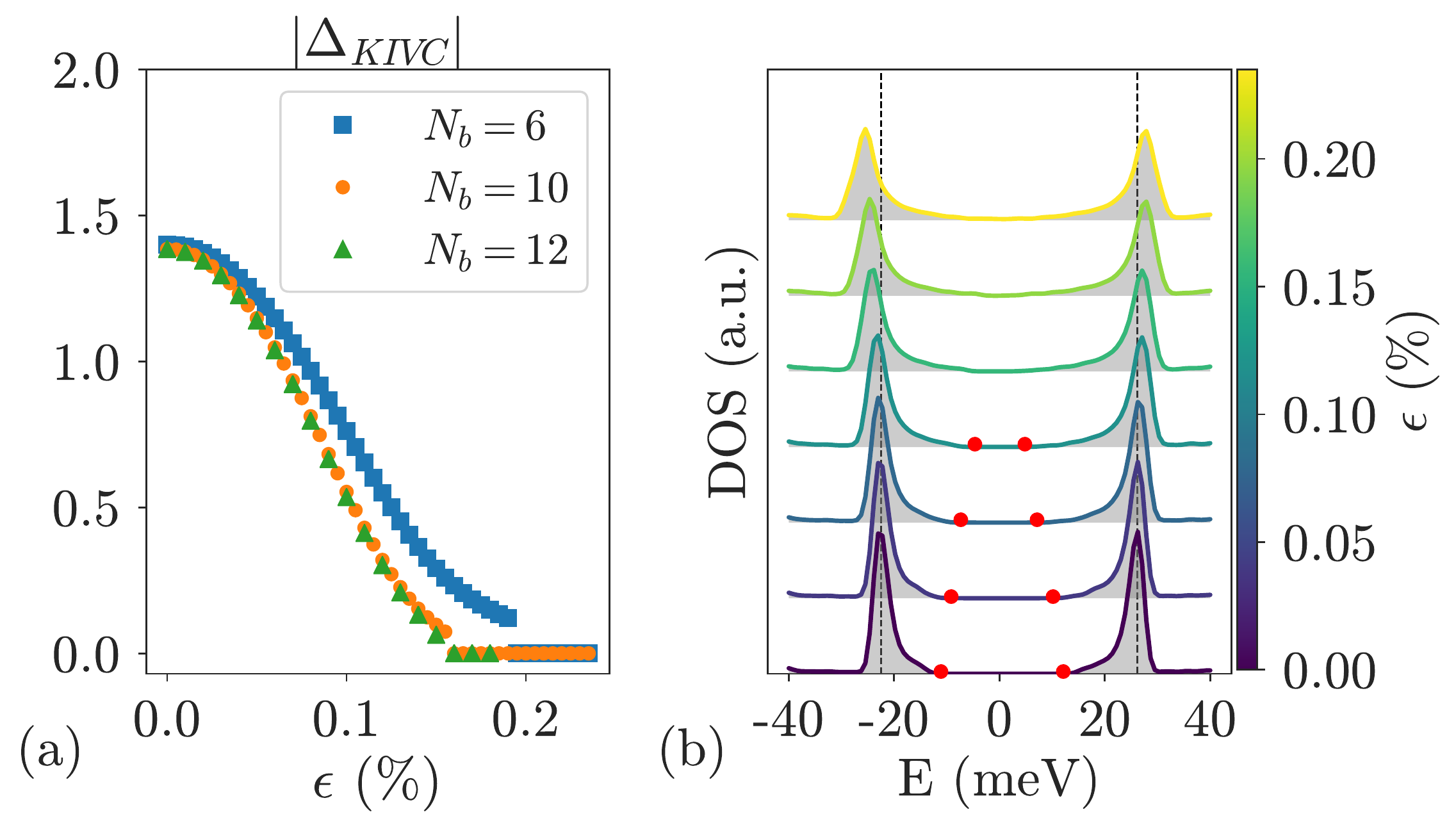}
    \caption{(a) KIVC order parameter $|\Delta_{\mathrm{KIVC}}| := \frac{1}{N}\sum_{\k} ||\m{P}_{\mathrm{IVC}}(\k)||$ at charge neutrality as a function of $\epsilon$, obtained with SCHF using $\theta = 1.05^\circ$, $\varepsilon_r = 10$ and $N_b=6,10$ or $12$ bands per spin and valley. The calculations were done on a $24\times 24$ momentum grid. (b) DOS of the SCHF band spectrum on a $36 \times 36$ momentum grid using $\theta = 1.05^\circ$, $\varepsilon_r = 10$ and $N_b=6$. The edges of the KIVC gap are indicated with red dots.
    }
    \label{fig:DOS}
\end{figure}

In Fig.~\ref{fig:DOS}(a) we plot the KIVC order parameter as a function of $\epsilon$. The order parameter is defined as $|\Delta_{\mathrm{KIVC}}| := \frac{1}{N}\sum_{\k} ||\m{P}_{\mathrm{IVC}}(\k)||$, where $N$ is the number of $\v{k}$ values and $\m{P}_{\mathrm{IVC}}2$ is the intervalley ($\tau \neq \tau'$) part of the KIVC correlation matrix $\left[\m{P}(\k)\right]_{(s,\tau,m),(s',\tau',n)} = \langle f^\dagger_{\k,s',\tau',n}f_{\k,s,\tau,m}\rangle$. We see that the transition occurs at $\epsilon_* \sim 0.19\%$ if we keep $N_b=6$ BM bands per spin and valley. By increasing $N_b$, $\epsilon_*$ shifts to slightly smaller values, and converges for $N_b=12$. Fig.~\ref{fig:DOS}(a) shows a discontinuity in $|\Delta_{\mathrm{KIVC}}|$, implying that the transition is first order. However, we also find that close to the transition, $|\Delta_{\mathrm{KIVC}}|$ decreases by a factor of $20$ (using $N_b=12$) compared to its value at $\epsilon = 0$. We therefore cannot exclude that the weakly first-order behavior is an artifact of HF.

\begin{figure}
    \centering
     \includegraphics[width=0.7\linewidth]{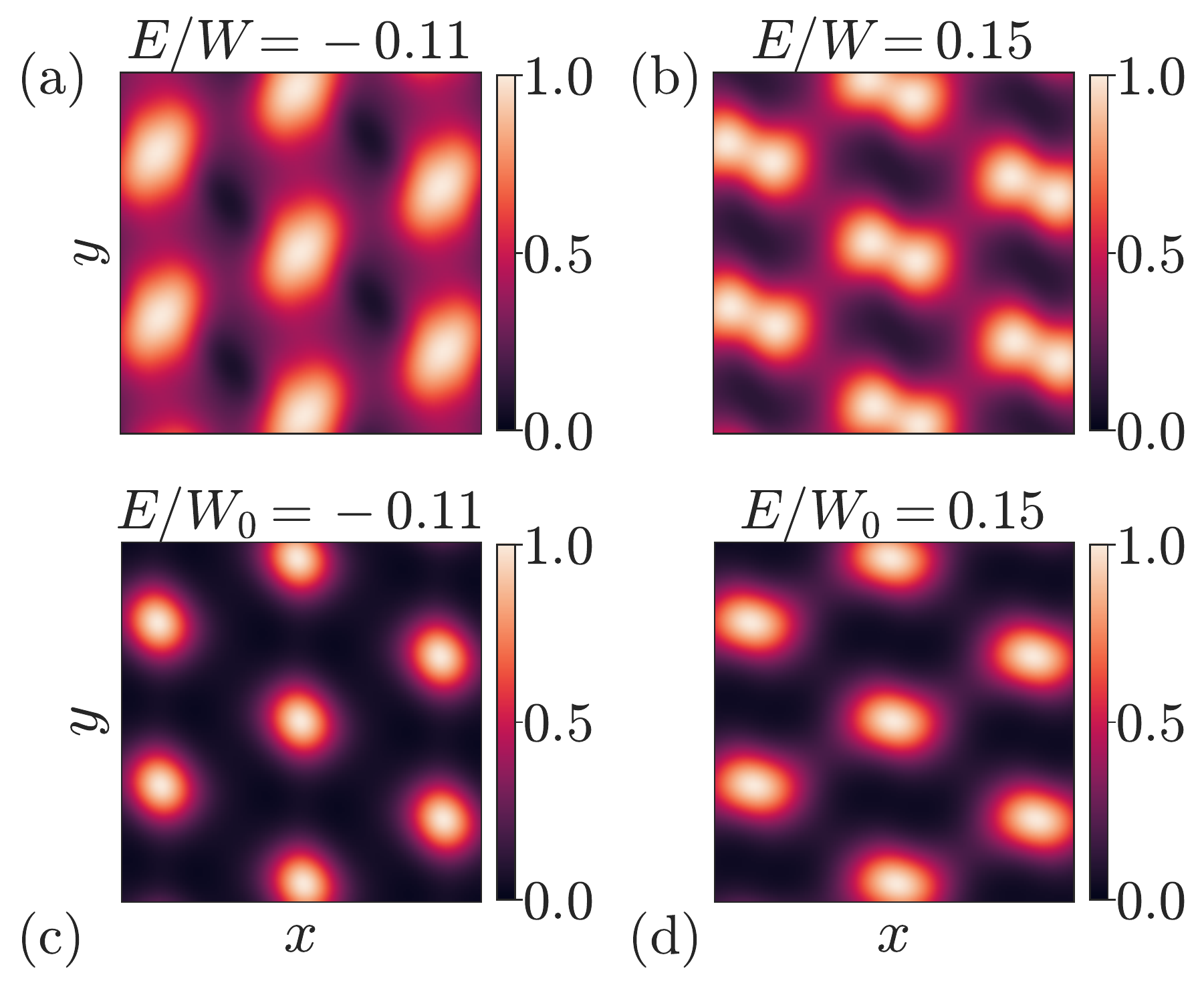}
    \caption{Normalized LDOS for $\theta=1.05^\circ$ and $\epsilon=0.22\%$. 
    (a)-(b) LDOS of the self-consistent SM (for $\varepsilon_r = 10$) at $E/W=-0.11$ and $E/W=0.15$, where $W\sim 65$ meV is the HF bandwidth. (c)-(d) LDOS of the BM ground state at $E/W_0=-0.11$ and $E/W_0=0.15$, where $W_0\sim 17$ meV is the BM bandwidth.}
    \label{fig:LDOS}
\end{figure}

Fig. \ref{fig:DOS}(b) shows the density of states (DOS) obtained in SCHF for different $\epsilon$, interpolating between the KIVC insulator and the SM. The dominant feature for both the KIVC and SM DOS is a pair of broad peaks separated by $\sim \SI{50}{\milli\electronvolt}$. In the KIVC phase, there is a finite window around the Fermi energy where the DOS is zero, which decreases with $\epsilon$ and vanishes at the transition. This is a subtle feature, however, making it hard to sharply distinguish the SM from the KIVC. A finer probe for the properties of the SM is the (layer-resolved) local DOS (LDOS) \cite{supplement}. In Fig. \ref{fig:LDOS}(a)-(b) we plot the LDOS of the SM at energies $E/W=-0.11$ and $E/W=0.15$, where $W$ is the HF bandwidth. The LDOS at the AA regions shows strong $C_{3z}$ breaking. This strong $C_{3z}$ breaking results from interactions, as it does not show up in the LDOS of the BM ground state at the same energy ratios $E/W_0=-0.11$ and $E/W_0=0.15$, where $W_0$ is the BM bandwidth (see Fig. \ref{fig:LDOS}(c)-(d) and \cite{supplement}). These properties of the HF LDOS agree with STM experiments \cite{CaltechSTM,RutgersSTM,ColumbiaSTM}. In particular, Ref. \cite{RutgersSTM} observed strong $C_{3z}$ breaking at the CNP, but not at $\nu = 4$. We calculated the LDOS at this filling, where the active bands are fully filled, and indeed found almost no reconstruction of the BM LDOS by interactions, and as a result no strong $C_{3z}$ breaking.
 
Finally, strain can be invoked to explain the degeneracies of the Landau fan near the CNP \cite{BiFu,PoSenthilC3} of the SM. At low densities quantum oscillations are governed by cyclotron orbits around the mini Dirac points, with two Dirac points for each of the four iso-spins. 
When  mirror symmetry ($C_{2x}$) ensures that the two Dirac points are equivalent, the resulting Landau fan will have the 8-fold degeneracy $\nu_\phi = \pm 4, \pm 12, \pm 20, \cdots$, which is  observed, for example, far from the magic angle. 
However, mirror symmetry is broken by strain: for example,  at $\epsilon = 0.22\%$ and $\varepsilon_r=10$, we find that the two  Dirac points in the same valley are separated in energy by $\Delta_D \sim 10$ meV.
For generic $B$, this halves the degeneracy, $\nu_\phi = 0, \pm 4, \pm 8, \pm 12, \cdots$, as observed in most magic-angle experiments \cite{PabloSC,Dean-Young}.
When $|\nu| \gtrsim 0.25$, the cyclotron orbits of the two Dirac points merge and form one connected orbit with a $2 \pi$-Berry phase. Because the resulting Landau fan $\nu_\phi = \pm 4, \pm 8, \pm 12, \cdots$ has the same 4-fold degeneracy as the $\Delta_D$-split Dirac points, the conclusion is the same.
However we note that some devices show a crossover from a low-$B$ 8-fold degeneracy to a high-$B$ 4-fold degeneracy (for example, at $B\sim 1$T in Ref.~\cite{YoungScreening}).
It may be that in devices where the strain configuration happens to produce a small $\Delta_D$, the mirror-breaking manifests in the terms which are linear in $B$.

\textit{DMRG at $\nu = -2$ --} While SCHF is a mean field approach, we may further  confirm the existence of a strain-induced transition using unbiased DMRG calculations. In Ref. \cite{KIVCpaper}, it was argued that in the absence of strain, the ground state of the interacting Hamiltonian $H$ at fillings $\nu=\pm 2$ is a spin polarized version of the KIVC state at neutrality. This claim was further substantiated by Refs. \cite{KangVafekPRL,ZhangJiang,TBGIV}.
Following the methods developed in Refs. \cite{DMRGpaper,KangVafek,tenpy}, here we use infinite DMRG to study $H$ compactified onto a infinitely long cylinder of circumference $L_y$ moire cells.
SCHF finds that  the ground state is perfectly spin polarized for $\epsilon \lesssim 0.2 \%$, so we accelerate our DMRG calculations by assuming full spin polarization of the narrow bands at $\nu = -2$, while keeping both valleys. \cite{supplement}. Projecting into  the narrow bands, our computational basis for the four remaining active bands consists of hybrid Wannier orbitals that are localized in the $x$-direction, but have a well-defined momentum $k_y = 2 \pi n / L_y$.

\begin{figure}
    \centering
    \includegraphics[width=0.98\linewidth]{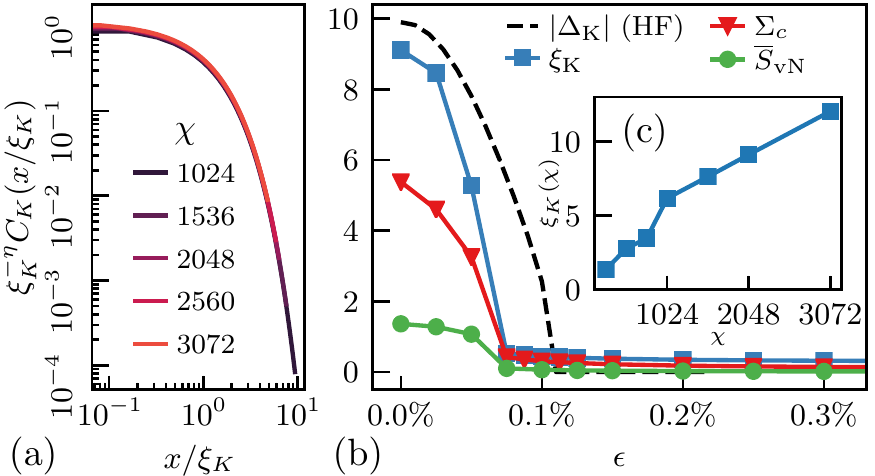}
    \caption{DMRG results at $\nu=-2$ (spin-polarized) at $\theta = 1.05^\circ$ and $\varepsilon_r=10$. (a) Scaling collapse of the KIVC correlator $C_K(x,\xi_K)$ at $\epsilon=0$. (b) Transition from KIVC to SM with strain. KIVC correlation length $\xi_K$, average entropy $\bar{S}_{vN}$, the DMRG KIVC correlator $\Sigma_C = 10\sum_x C_K(x)$ (scaled for visibility), and the HF KIVC correlator $|\Delta_{\text{KIVC}}|$ as a function of $\epsilon$. (c) Scaling of $\xi_K$ with bond dimension at $\epsilon=0$. DMRG parameters: $L_y = 6$, $\Phi_y = 0$, $\chi\approx 2048$ for (b), and the Hamiltonian, Eq. \eqref{eq:IBM}, is represented to accuracy better than \SI{0.1}{\milli\electronvolt}. All quantities are defined in the text.
    }
    \label{fig:DMRG}
\end{figure}

The ground state of the unstrained model at $\nu=-2$ is expected to have KIVC order, and thus to spontaneously break the $U(1)$ valley symmetry. The Hohenberg-Mermin-Wagner (HMW) theorem, however, forbids such continuous symmetry breaking on the quasi-1D cylinder geometry used by DMRG \cite{Hohenberg,MerminWagner}. Instead, the KIVC phase will manifest as algebraic long-range order \cite{DMRGSkyrmionSC} $C_K(x) := \langle \Delta_K^+(x)\Delta_K^-(0)\rangle \sim x^{-\eta(L_y)}$, where $\Delta_{K}^\pm(x)$ are operators at position $x$ which have valley charge $\pm 2$ and satisfy $\mathcal{T}'^{-1}O_K^\pm(x) \mathcal{T}' = O_K^{\mp}(x)$ \cite{supplement}. The exponent $\eta(L_y)$ depends on the circumference, and satisfies $\eta(\infty) = 0$. An additional complication for identifying the KIVC phase using DMRG is that at any finite DMRG bond dimension $\chi$ (i.e., numerical accuracy), the ground state has exponentially decaying  correlations. This complication can be overcome by using ``finite entanglement scaling'' \cite{PollmannTurner, Tagliacozzo,Kjall} to characterize algebraic order via a scaling collapse as $\chi \to \infty$. Denoting the finite-$\chi$ induced correlation length as $\xi_K$ [Fig. 4(c)], the KIVC correlator can be written as a general function $C_K(x,\xi_K)$. In the KIVC phase, we expect this function to satisfy the scaling relation $C_K(x,\xi_K) = \xi_K^{-\eta}C_K(x/\xi_K,1)$, which allows us to perform a scaling collapse of the data obtained at different $\chi$. In Fig. \ref{fig:DMRG}(a), we find an excellent data collapse for $\chi$ ranging between $1024$ and $3072$, from which we conclude that DMRG indeed finds a KIVC ground state. Note that we find a very small exponent $\eta(6) \sim 0.06$ \cite{supplement}, so there is no regime of algebraic decay clearly visible in Fig. \ref{fig:DMRG}(a).

Fig. \ref{fig:DMRG} (b) shows the effect of adding strain. Both the correlation length $\xi_K$ and summed correlator $\Sigma_C := \sum_x C_K(x)$ measure the amount of KIVC correlations in the ground state. They are both order one for small strain, and decrease monotonically with $\epsilon$. For $\epsilon \gtrsim 0.07\%$, however, $\xi_K$ and $\Sigma_C$ plateau at a small value, indicating that the algebraic KIVC order is destroyed. For strain values larger than $\sim 0.07\%$, we find no evidence for symmetry breaking in the DMRG ground state. In particular, we have verified that DMRG does not double the unit cell, which excludes the stripe phase discussed previously for single-valley models \cite{KangVafek,DMRGpaper}. The absence of symmetry breaking in DMRG is consistent with HF, where we find a symmetric SM at large $\epsilon$ \cite{supplement}. Fig 4(b) plots the SCHF order parameter $|\Delta_{\mathrm{KIVC}}|$, which shows a transition from the KIVC state to the SM at a strain value $\epsilon\sim 0.1 \%$, close to where the algebraic KIVC order disappears in DMRG. To confirm that the large strain phase found with DMRG is the same SM obtained in SCHF, we compute the averaged single particle entropy $\bar{S}_{\mathrm{vN}}:=-\frac{1}{N}\sum_\k \text{tr}\left(\m{P}(\k)\ln \m{P}(\k) \right)$. This quantity is zero iff the DMRG ground state is a Slater determinant. Fig 4(b) shows that $\bar{S}_{\mathrm{vN}}$ is negligibly small at $\epsilon\gtrsim 0.07\%$ (at smaller $\epsilon$, HMW implies the KIVC state cannot be a symmetry breaking Slater determinant in DMRG, so $\bar{S}_{\mathrm{vN}}$ is order unity). It thus follows that (i) SCHF and DMRG agree closely for all strain, and are essentially identical at large $\epsilon$ and, (ii) the transition in DMRG is indeed from the KIVC state to the SM.

\textit{Discussion --} The results presented in this work show that strain is likely responsible for the semi-metallic behavior and strong $C_{3z}$ breaking observed at the CNP of most TBG devices (for related discussions of the CNP physics, see Refs. \cite{Savary,OchoaStrain}). $C_{3z}$ breaking has also been observed in TBG near $\nu = -2$ \cite{PabloNematic}, and was discussed in various theoretical contexts in Refs.  \cite{FernandesVenderbos,FernandesNematicSC,ChubukovNematic,KoziiNematic}. From our DMRG and SCHF results, we found that TBG couples strongly to strain both at $\nu = 0$ and $\nu=-2$. Two important questions that follow from this are (i) whether the strong coupling to strain persists to $\nu = -2 - \delta$ with $\delta\sim 0.1-0.9$ (where nematicity was observed in experiment \cite{PabloNematic}), and (ii) whether strain is important for superconductivity. Our findings also invigorate the question about the origin of the insulating behavior consistently  observed at $\nu = -2$, as we find that within the model studied here, strain drives the KIVC - SM transition at roughly the same $\epsilon$ for both $\nu = 0$ and $\nu = -2$. One possibility is that band structure effects we have neglected, such as lattice relaxation \cite{NamKoshino,CarrKaxiras} or non-local inter-layer tunneling \cite{CarrKaxiras,XieMacDonaldHall} stabilize the insulators at $\nu = \pm 2$ at larger strain values.

\textit{Acknowledgements --} We want to thank Eslam Khalaf, Shubhayu Chatterjee and Ashvin Vishwanath for helpful discussions. NB would like to thank Glenn Wagner and Yves Kwan for useful feedback on an earlier version of this manuscript. NB was supported by a fellowship of the Research Foundation Flanders (FWO) under contract no. 1287321N. DEP was funded by the Gordon and Betty Moore Foundation's EPiQS Initiative, Grant GBMF8683. MPZ was supported by the Director, Office of Science, Office of Basic Energy Sciences, Materials Sciences and Engineering Division of the U.S. Department of Energy under contract no. DE-AC02-05-CH11231 (van der Waals heterostructures program, KCWF16). 
JH was funded by the U.S. Department of Energy, Office of Science, Office of Basic Energy Sciences, Materials Sciences and Engineering Division under Contract No. DE-AC02-05- CH11231 through the Scientific Discovery through Advanced Computing (SciDAC) program (KC23DAC Topological and Correlated Matter via Tensor Networks and Quantum Monte Carlo).
This research used the Savio computational cluster resource provided by the Berkeley Research Computing program at the University of California, Berkeley (supported by the UC Berkeley Chancellor, Vice Chancellor for Research, and Chief Information Officer).

\bibliography{references}

\clearpage
\onecolumngrid
\appendix

\begin{center}
    \textbf{SUPPLEMENTARY MATERIAL FOR `STRAIN-INDUCED QUANTUM PHASE TRANSITIONS IN MAGIC ANGLE GRAPHENE'}
\end{center}

\section{Properties of the strained BM model and the self-consistent semi-metal at neutrality}

In this appendix, we discuss additional properties of both the non-interacting BM model in the presence of non-zero strain, and the self-consistent semi-metal obtained in HF at the CNP for sufficiently large $\epsilon$.

\begin{figure}[h]
    \centering
    \includegraphics[scale=0.4]{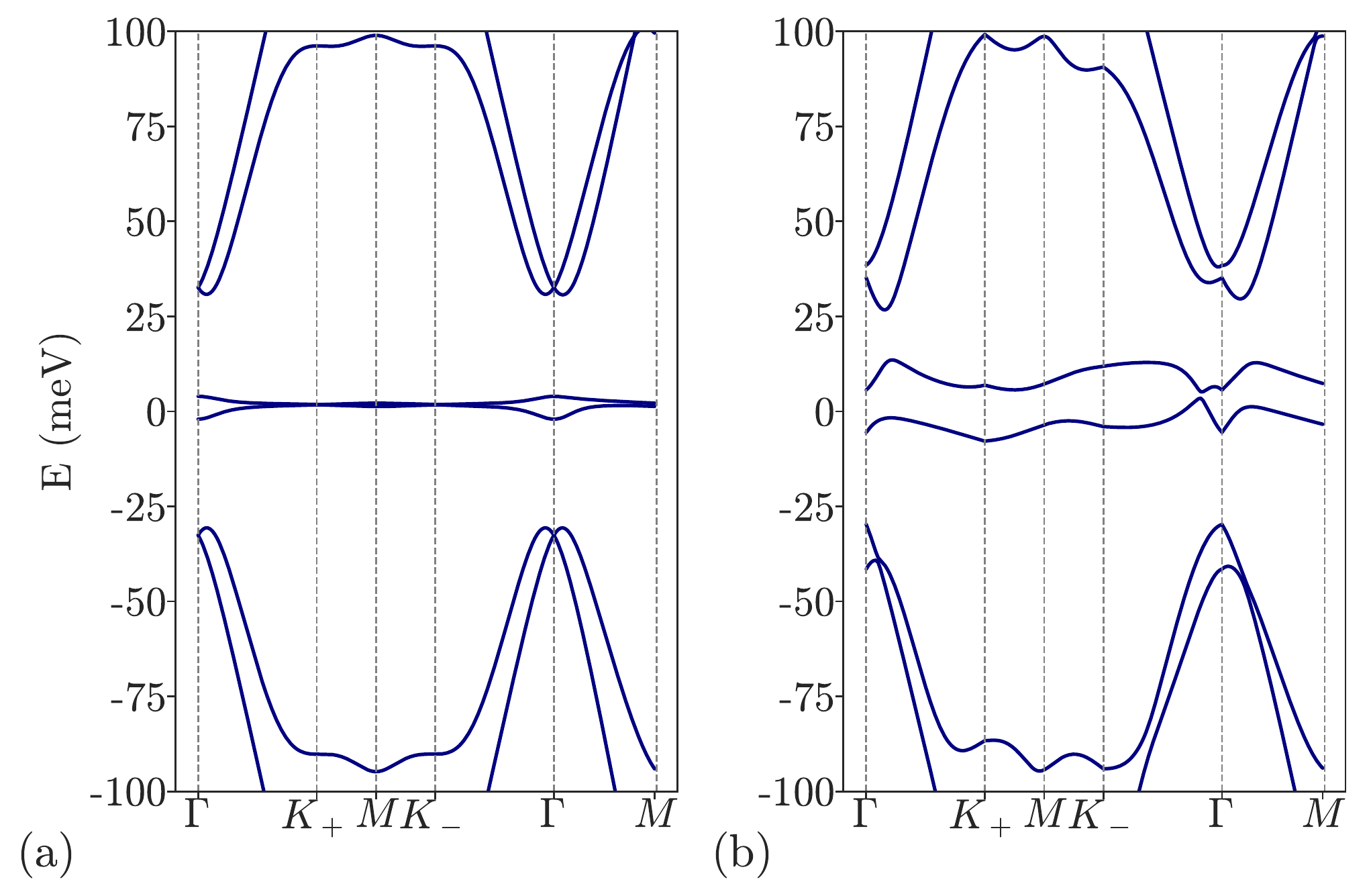}
    \caption{Band spectrum of the single-valley BM model with twist angle $\theta=1.05^\circ$ along a cut through the mini-BZ. (a) Original BM model with $\epsilon = 0$. (b) Strained BM model with $\epsilon = 0.3\%$.}
    \label{fig:bandspectra}
\end{figure}

Fig. \ref{fig:bandspectra} shows the single-valley BM band spectrum along a cut through the 2D mBZ, using $\theta = 1.05^\circ$, and both $\epsilon = 0$ and $\epsilon = 0.3\%$. At finite strain, the bandwidth of the two active bands is much larger than the bandwidth at $\epsilon =0$. Importantly, for $\epsilon = 0.3\%$ there is still a sizable gap between the active bands and the remote bands. This is especially important for our DMRG simulations, which work with an interacting Hamiltonian projected into the active bands only.

\begin{figure}
    \centering
    \includegraphics[scale=0.27]{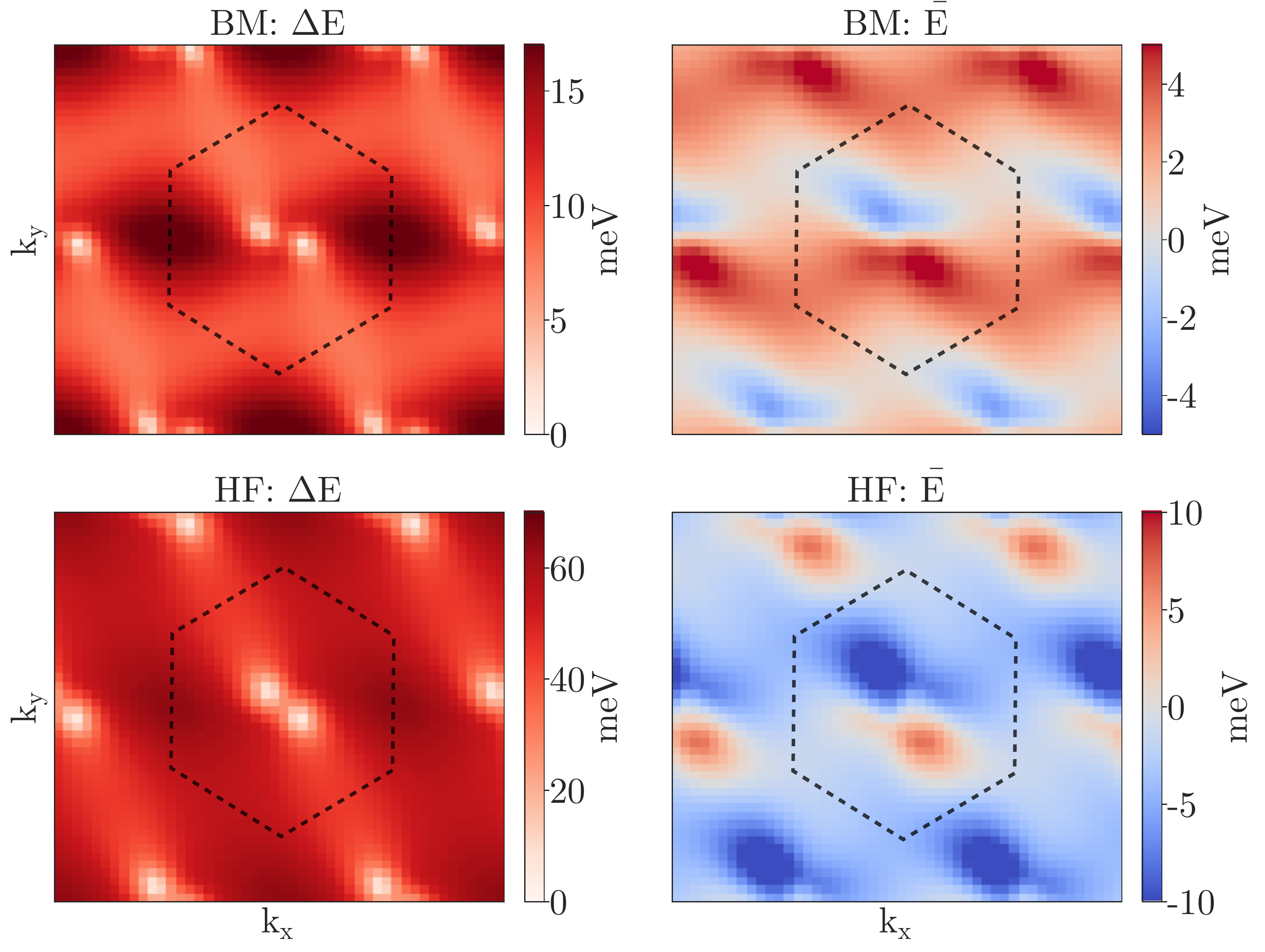}
    \caption{BM and HF band energies in the $\tau = +$ valley using $\theta=1.05^\circ$ and $\epsilon = 0.22\%$. A coordinate transformation is performed in momentum space such that even with non-zero strain, the mBZ is a regular hexagon, indicated by the dashed lines. (a)-(b) Energy difference $\Delta E$ and average energy $\bar{E}$ of the two active bands of the BM Hamiltonian. (c)-(d) Energy difference $\Delta E$ and average energy $\bar{E}$ of the two active HF bands of the self-consistent SM. A dielectric constant $\varepsilon_r = 10$, a $24\times 24$ momentum grid and $N_b=6$ bands per spin and valley were used.}
    \label{fig:BS}
\end{figure}

In Fig. \ref{fig:BS}(a)-(b), we plot the difference and the average of the two active band energies of the single-valley BM Hamiltonian in the entire mBZ, using a non-zero strain $\epsilon = 0.22\%$. Note that we have performed a coordinate transformation in momentum space such that the mBZ is a regular hexagon. From the difference in energies, one can clearly identify the position of the two Dirac cones, which as mentioned in the main text are in the vicinity of the $\Gamma$-point. 

In Fig. \ref{fig:BS}(c)-(d), we plot the energy difference and average of the two active bands in the mean-field band spectrum of the self-consistent SM in the $\tau=+$ valley at charge neutrality. The self-consistent SM was obtained using a strain value $\epsilon=0.22\%$, at which the KIVC order is destroyed and the SM is the lowest-energy state. Similarly to the non-interacting BM Hamiltonian, the Dirac cones of the self-consistent SM are located near $\Gamma$. Away from these Dirac points, however, the two active bands in the $\tau=+$ valley are now separated by an energy difference of roughly $50$ meV (using $\varepsilon_r = 10$), which is larger by a factor of $5$ compared to the energy separation in the non-interacting BM model at the same value of $\epsilon$. 

\begin{figure}
    \centering
    \includegraphics[scale=0.37]{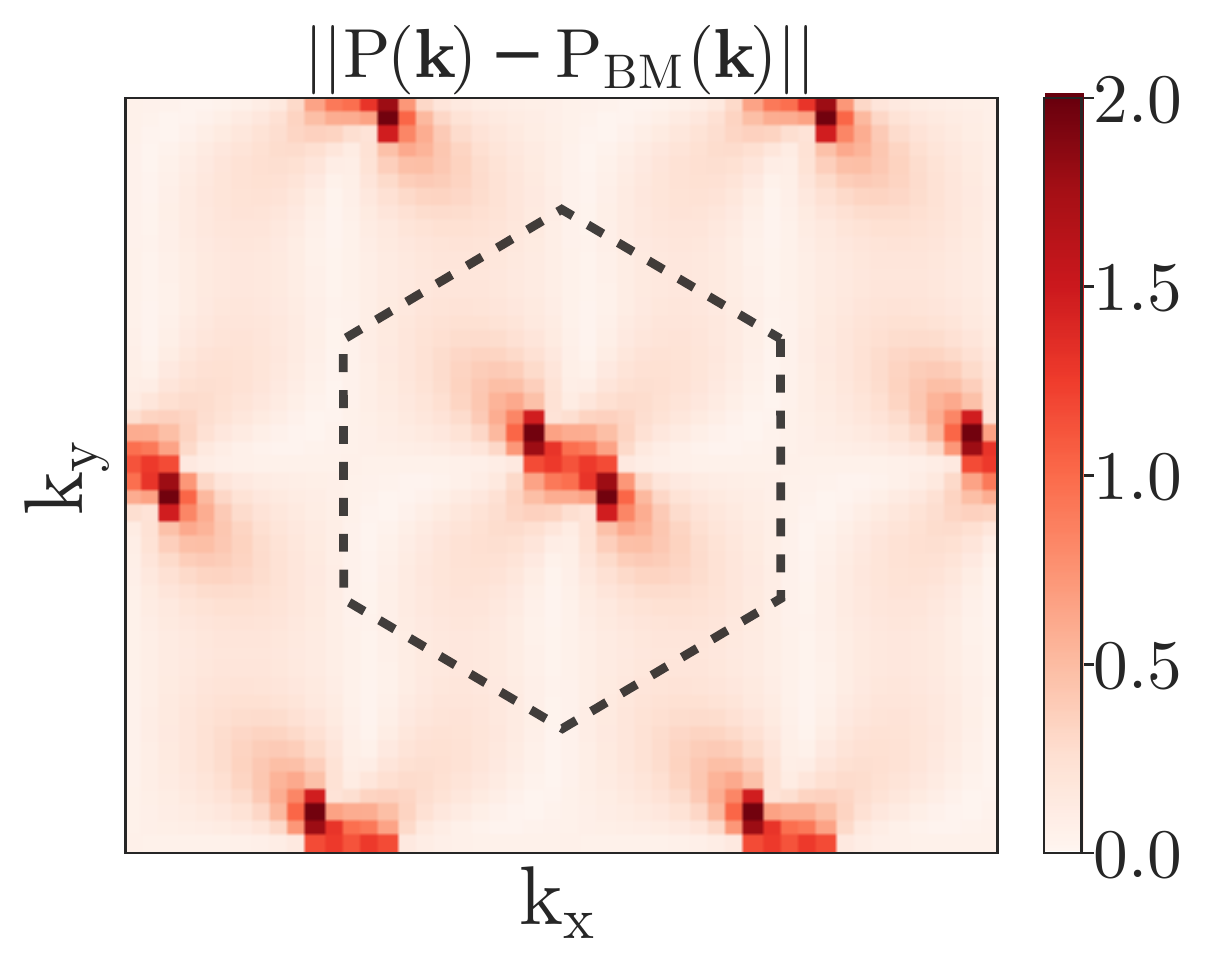}
    \caption{Norm of the difference between the correlation matrix $\m{P}(\k)$ of the self-consistent SM and the correlation matrix $\m{P}_{\mathrm{BM}}(\k)$ of the ground state of the non-interacting BM Hamiltonian at charge neutrality, using twist angle $\theta=1.05^\circ$ and strain $\epsilon=0.22\%$. A coordinate transformation is performed in momentum space such that the mBZ is a regular hexagon, indicated by the dashed lines. The self-consistent SM is obtained using $\varepsilon_r=10$ and $N_b=6$ bands per spin and valley.}
    \label{fig:Pdiff}
\end{figure}

To quantify how different the BM ground state is from the self-consistent SM at charge neutrality, we plot the Frobenius norm of $\m{P}(\k)-\m{P}_{\mathrm{BM}}(\k)$ in Fig. \ref{fig:Pdiff}. As in the main text, $\left[\m{P}(\k)\right]_{(s',\tau',m),(s,\tau,n)} = \langle f^\dagger_{\k,s,\tau,n}f_{\k,s',\tau',m}\rangle$ is the correlation matrix of the self-consistent Slater determinant with the lowest energy, which is the SM for the strain value $\epsilon=0.22\%$ used in Fig. \ref{fig:Pdiff}. $\left[\m{P}_{\mathrm{BM}}(\k)\right]_{(s',\tau',m),(s,\tau,n)} = \delta_{s,s'}\delta_{\tau,\tau'}\delta_{m,n}\Theta(\varepsilon_{\k,\tau,n})$, with $\Theta(x)$ the Heaviside step function, is the correlation matrix of ground state of the non-interacting BM Hamiltonian at charge neutrality. From Fig. \ref{fig:Pdiff}, we see that $|| \m{P}(\k)-\m{P}_{\mathrm{BM}}(\k)||$ is equal to $\sim 0.1$ almost everywhere in the mBZ, except close to the $\Gamma$ point, where it becomes of order one. This shows that the self-consistent SM has significant overlap with the BM ground state in most of the mBZ already at small strain values.

The discrepancy between the BM ground state and the self-consistent SM in a small region near the $\Gamma$ point is responsible for the differences in the LDOS discussed in the main text. Here, we further elaborate on this point. We define the energy and layer-resolved LDOS as

\begin{equation}
    \rho_\ell(E,\r) = \frac{1}{A_{mBZ}}\sum_{s,\tau,n} \int \mathrm{d}^2\k\; \delta(E - \varepsilon_{\k,s,\tau,n}) \sum_{\sigma=A,B}|\psi^{\ell,\sigma}_{\k,s,\tau,n}(\r)|^2\, ,
\end{equation}
where $\ell = \pm$ denotes the graphene layers, $A_{mBZ}$ is the area of the mBZ, $\varepsilon_{\k,s,\tau,n}$ are the single-particle energies of the mean-field band spectrum, and $\psi^{\ell,\sigma}_{\k,s,\tau,n}(\r)$ are the corresponding single-particle wavefunctions. In practice, we calculate $\rho_\ell(E,\r)$ by replacing the momentum integral by a discrete sum, and the delta-function by a Gaussian with a standard deviation of $\sim 0.5$ meV for the self-consistent SM, and $\sim 0.2$ meV for the BM ground state. 

\begin{figure}
    \centering
    \includegraphics[scale=0.28]{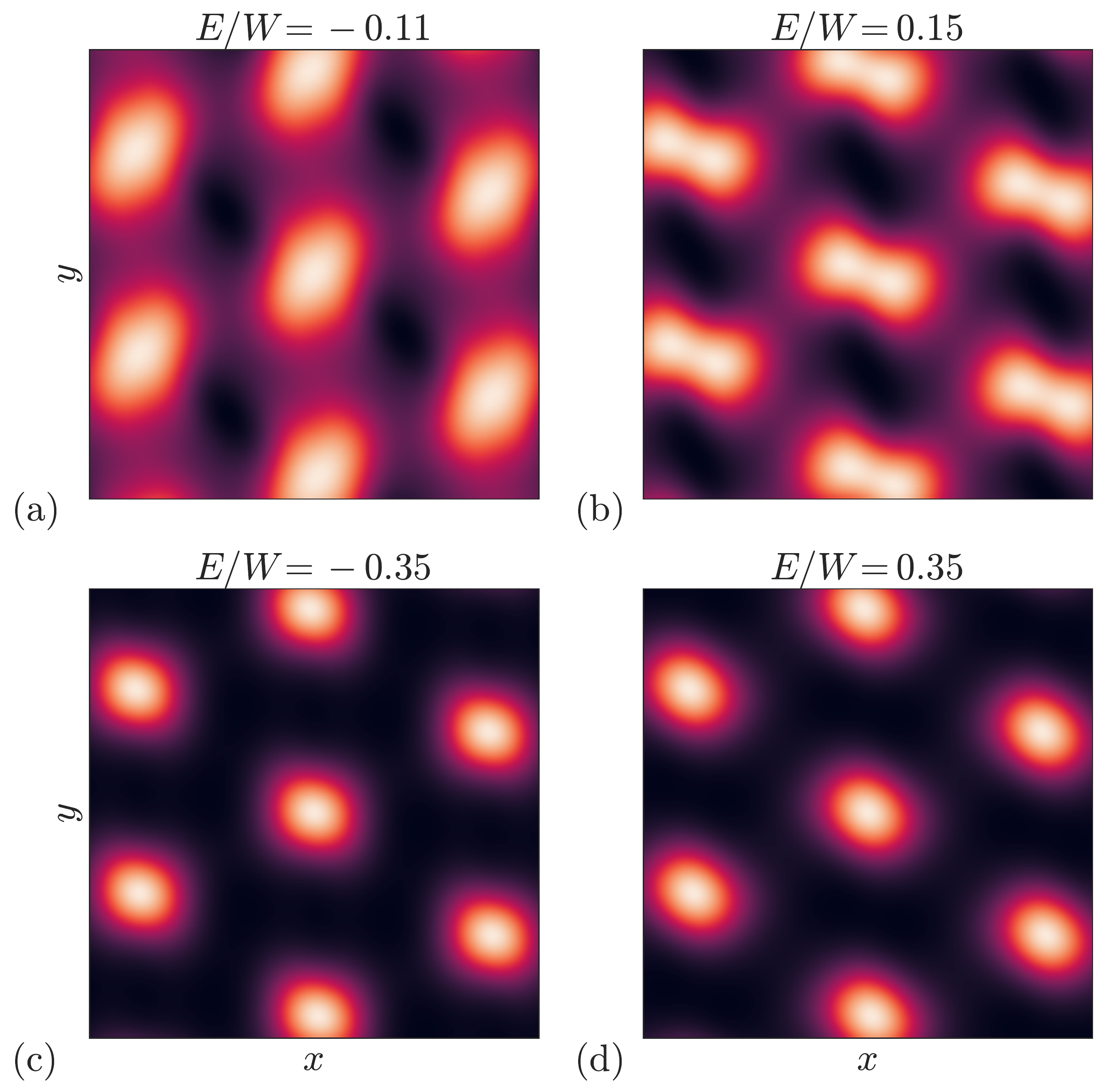}
    \caption{(a)-(d) Local density of states $\rho_{+}(E,\r)$ on the top layer (in arbitrary units) of the self-consistent SM obtained in HF. An overall energy constant is fixed by requiring that the HF single-particle energies $\varepsilon_{\k,n}$ satisfy $\sum_\k\sum_n \varepsilon_{\k,n} = 0$. The results were obtained on a $24\times 24$ momentum grid using $\theta=1.05^\circ$, $\epsilon = 0.22\%$, $\varepsilon_r=10$, and six bands per spin and valley.}
    \label{fig:LDOS_HF}
\end{figure}

In Fig. \ref{fig:LDOS_HF}, we plot $\rho_+(\r,E)$ for $E/W=-0.35, -0.11, 0.15$ and $0.35$, where $W\sim 65$ meV is the HF bandwidth (an overall energy constant is fixed by imposing that $\sum_{\k,n}\varepsilon_{\k,n} =0)$. In Fig. \ref{fig:LDOS_HF}(a)-(b), we see a very clear $C_{3z}$-breaking in the LDOS at energies $E/W=-0.11$ and $E/W=0.15$, which respectively correspond to $E \sim -7$ meV and $E \sim 10$ meV and thus lie outside the broad peaks in the DOS $\rho(E)=\int\mathrm{d}^2\r \sum_\ell \rho_\ell(E,\r)$ (see Fig. \ref{fig:DOS} in the main text). At larger energies $E/W=\pm 0.35$, which correspond to values $E\sim\pm 20$ meV inside the broad peaks in the DOS, the $C_{3z}$ breaking is also present, but is much less pronounced. At $E/W=-0.11$ and $E/W=0.15$, the local charge distributions at the AA regions are clearly elongated in one direction. Contrary to what one might expect, these strongly $C_{3z}$-breaking charge distributions are not simply a consequence of strain, but instead rely on the Coulomb interaction. In particular, we find that for any value of $E$ inside the active bands, $\rho_+(E,\r)$ obtained from the non-interacting BM ground state does not show the same clear asymmetric charge distributions at the AA regions as does the self-consistent SM (for an example of the BM LDOS at two representative energies, see Fig. \ref{fig:DOS} in the main text). This implies that interactions are necessary to reconstruct the BM LDOS in order to obtain strong $C_{3z}$ breaking.

\section{Hartree-Fock at $\nu = -2$}

\begin{figure}
    \centering
    \includegraphics[scale=0.38]{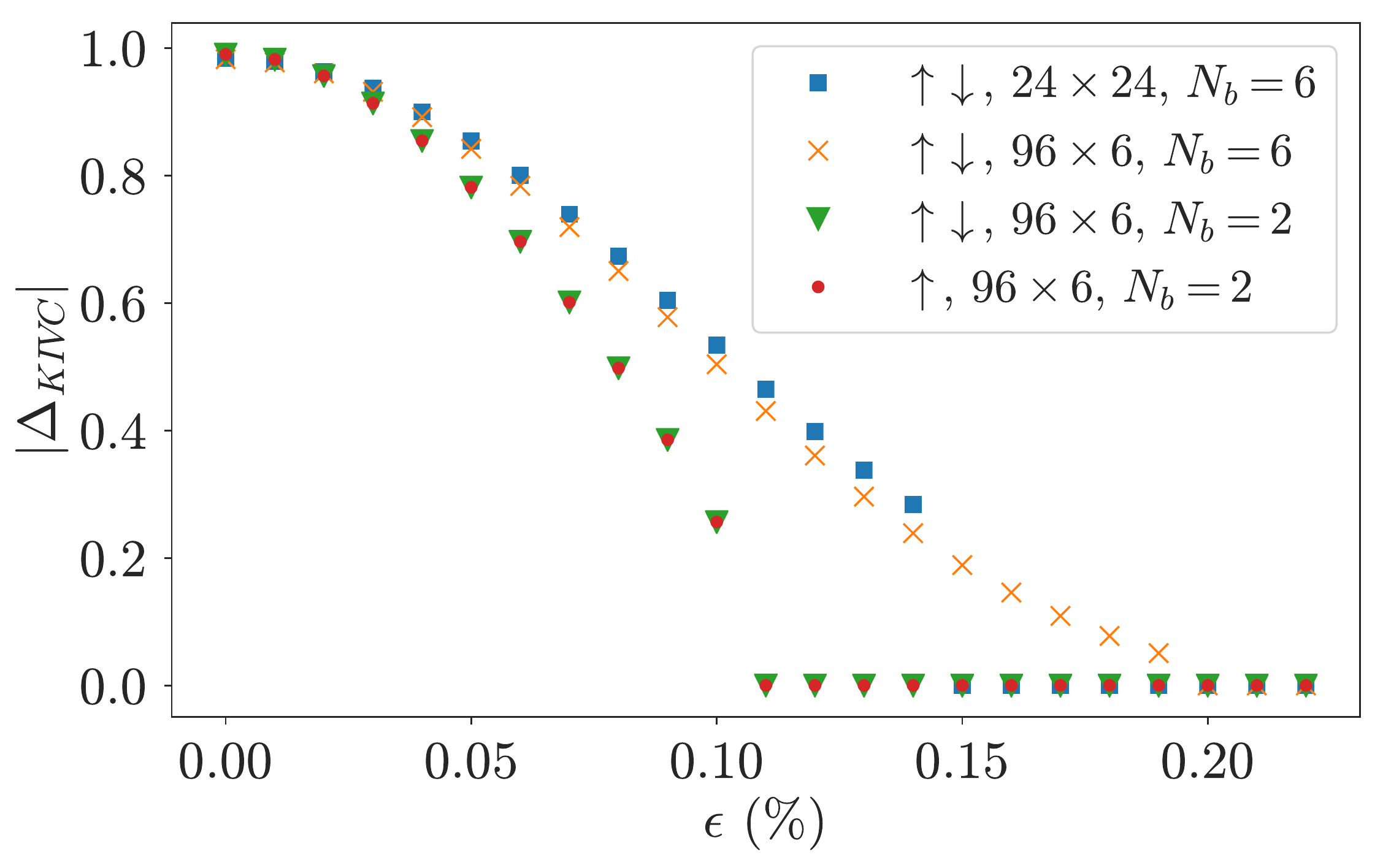}
    \caption{Hartree-Fock KIVC order parameter as a function of strain $\epsilon$ at $\nu = -2$. $\uparrow\downarrow$ means that both spin flavors were used, while $\uparrow$ means that the active bands were taken to be spin polarized. The momentum grids used were of sizes $24\times 24$ and $96\times 6$ (with the smallest direction being the $y$-direction). $N_b$ is the number of bands kept per spin and valley.
    }
    \label{fig:HFnu2}
\end{figure}

In Fig. \ref{fig:HFnu2} we show the SCHF results at filling $\nu = -2$. In particular, we plot the KIVC order parameter $|\Delta_{\mathrm{KIVC}}|$ as a function of $\epsilon$. We have performed four SCHF calculations, each with a different set-up. The first SCHF calculation was done using both spin flavors on a $24\times 24$ momentum grid, keeping $N_b=6$ bands per spin and valley. The second calculation was done on a rectangular $96\times 6$ momentum gird, also with two spin flavors and six bands. The third SCHF calculation was again done on the same rectangular grid with two spin flavors, but now keeping only $N_b=2$ bands per spin and valley. In the fourth and final SCHF calculation we reduced the number of spin components in the active bands from two to one (again working on the same rectangular grid and using $N_b=2$). Importantly, even though we keep only one spin flavor for the active bands, the remote bands retain two spin flavors. This shows up in our Hamiltonian via the HF contribution of the remote bands to the free fermion part $h(\k)$ of $H$ [Eq. \eqref{eq:IBM}].

From Fig. \ref{fig:HFnu2} we see that going from the square to the rectangular momentum grid stabilizes the KIVC state over larger strain values. On reducing the number of bands $N_b$ from six to two, however, the strain interval over which we find KIVC order becomes significantly smaller. For all SCHF calculations where we keep both spin flavors, we find that the spin polarization $P_s := \frac{1}{N}\sum_\k \text{tr}\left(\m{P}(\k)s_z\right)$, with $s_z$ the Pauli-z matrix acting on spin indices, is independent of $\epsilon$ and retains its value $P_s =2$. This is because the filled active bands are completely spin polarized by the exchange interaction. As a result, we see that the SCHF calculation on the rectangular grid with only one spin component for the active bands produces results that are indistinguishable from the results obtained for the complete model with both spin flavors. This justifies doing DMRG on the model with spin polarized active bands. We also see that the transition from the KIVC to the SM on the rectangular grid with $N_b = 2$ happens near $\epsilon \sim 0.1\%$, which is very close to the value where DMRG puts the phase transition (see main text). 

\section{Details of the DMRG Calcuations}

\begin{figure}
    \centering
    \includegraphics[width=0.5\linewidth]{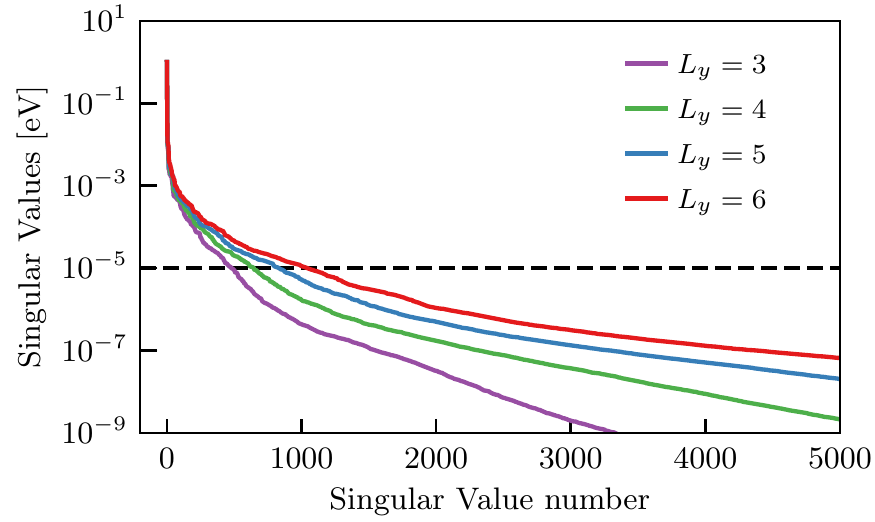}
    \caption{The MPO singular values \cite{parker2019local} for the Hamiltonian as a function of $L_y$. The dashed line represents our truncation level.}
    \label{fig:MPO_SV_spectra}
\end{figure}

Our DMRG calculations follow the method described in \cite{DMRGpaper}. In brief, we start with a Bistrizer-MacDonald-like continuum model for TBG. We perform 1D hyrid Wannier localization which gives states that are localized in $x$ and periodic along $y$. These states form the computational basis and have corresponding creation operators $c^\dagger_{x,k_y,\sigma,\tau}$ where $x \in \N$ indexes the $x$ position, $k_y = 2\pi n/L_y$ runs over $L_y$ momentum cuts through the mBZ, $\sigma = \pm 1$ labels sublattice, and $\tau = \pm$ labels the $K$ and $K'$ valleys. Each unit cell has $4L_y$ tensors, and respects $U(1)$ charge and valley symmetry. We then add Coulomb interactions as proscribed in Eq. \eqref{eq:IBM}, with a cutoff of $6$ moir\'e unit cells. The matrix product operator (MPO) for this Hamiltonian has bond dimension $D \approx 40,000$ at $L_y =6$ --- large, but unsurprising since this is a 2D model with a large unit cell, and long-range interactions. We use the MPO compression procedure of \cite{parker2019local} to reduce this to $D \approx 1100$ while retaining a precision of \SI{0.01}{\milli\electronvolt}, as shown in Fig. \ref{fig:MPO_SV_spectra}. Our DMRG is performed using the \texttt{TeNPy} library \cite{tenpy}, written by one of us.

\subsection{KIVC order parameter and correlation lengths with DMRG}

This section details the KIVC order parameter used for DMRG. As mentioned in the main text, the KIVC phase breaks both valley charge conservation symmetry $e^{i \alpha \tau^z}$ as well as spinless time-reversal symmetry $\mathcal{T} = \tau^x K$ ($K$ is complex conjugation), but preserves the product $\mathcal{T}' = e^{i\pi \tau^z/2} \mathcal{T} = \tau^y \mathcal{T}$. In 2D, the KIVC phase can be detected by the order parameters $\int d^{2}\v{k} \; O^\pm_K(\v{k})$, where $O_K^\pm(\v{k}) = \sigma^y \tau^{\pm}(\v{k})$. Here we have introduced the notation $\sigma^i\tau^j(\v{k}) := c^\dagger_\k \sigma^i\tau^jc_\k$, and $\tau^\pm = \tau^x\pm i \tau^y$. The Pauli matrices $\sigma^i$ act on the orbital indices of the two hybrid Wannier states in the same valley. The creation operators $c^\dagger_{\k,\sigma,\tau}$ are in turn defined as $c^\dagger_{k_x,k_y,\sigma,\tau} := \sum_{x \in \N}  e^{-i x (k_x + k_y/2)} c^\dagger_{x,k_y,\sigma,\tau}$ (in units where $k_x \in [0,2\pi)$ ).

It is important to distinguish the KIVC phase from the time-reversal intervalley coherent (TIVC) phase, whose order parameters are $O_T^\pm(\v{k}) = \sigma^x \tau^{\pm}(\v{k})$. The following table shows how these operators behave under conjugation by symmetries: $O \to U^{-1} O U$.
\begin{center}
\begin{tabular}{ l c c }
\toprule
Symmetry & $O^\pm_{T}(\v{k})$ & $O^\pm_K$\\[0.3em] \colrule
$\mathcal{T}$ & $O^\mp_T(-\v{k})$ & $-O^\mp_K(-\v{k})$\\[0.3em]
$\mathcal{T}'$ & $-O^\mp_T(-\v{k})$ & $O^\mp_K(-\v{k})$\\[0.3em]
$e^{i\alpha\tau_z}$ & $e^{\pm i2\alpha}O^\pm_T(\v{k})$ & $e^{\pm i2\alpha}O^\pm_K(\v{k})$\\[0.3em]
\botrule
\end{tabular}
\end{center}
From this table it follows that $O^x_K(\k) := [O^+_K(\k) + O^-_K(\k)]/2$, averaged over the whole BZ, vanishes for $\mathcal{T}$-symmetric phases but is generically non-zero for $\mathcal{T}'$-symmetric states --- and therefore distinguishes between the KIVC and TIVC states.

\begin{figure}
    \includegraphics[width=0.9\linewidth]{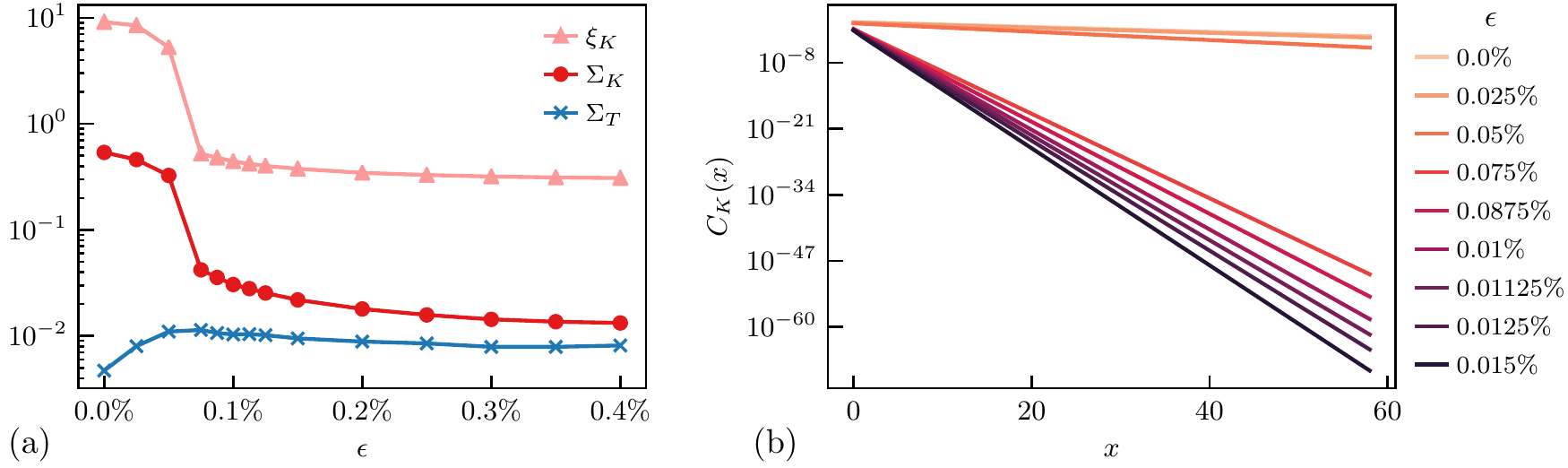}
	\caption{(a) The correlation length in the KIVC sector $\xi_K$, as well as the summed KIVC and TIVC correlators $\Sigma_K = \sum_x C_K(x)$ and $\Sigma_T = \sum_x C_T(x)$ plotted as a function of strain $\epsilon$. Note that $\xi_K$ diverges with $\chi$ in the KIVC phase, as shown in Fig 4(c).
	One can see that $\Sigma_K$ tracks $\xi_K$ closely, but $\Sigma_T$ does not, and that $\Sigma_T \ll \Sigma_K$. (b) The correlator $C_K(x)$ at a range of strain values. The transition is clearly visible. Parameters: $\nu = -2, L_y = 6,\Phi_y = 0, \chi = 2048$.}
	\label{fig:KIVC_vs_TIVC}
\end{figure}

DMRG, as noted in the main text, uses a quasi-1D geometry which cannot support the spontaneously broken $U(1)$ valley symmetry that accompanies the KIVC phase. However, two-point functions of any order parameter with the correct symmetry quantum numbers will exhibit long-range order with a diverging correlation length. To that end, we consider
\begin{equation}
    C_K(x) = \braket{\Psi| \left(\sum_{k_y = \pm k_{y0}} O_{K}^+(x,k_y)\right) \left(\sum_{k_y' = \pm k_{y0}} O_{K}^-(0,k_y') \right) |\Psi}.
    \label{eq:DMRG_KIVC_order_param}
\end{equation}
where 
\begin{equation}
O_K^\pm(x,k_y) = [\sigma^y\tau^\pm](x,k_y) = i c^\dagger_{x,k_y,\sigma = 1,\tau=\pm 1} c_{x,k_y,\sigma=-1,\tau=\mp 1} -  i c^\dagger_{x,k_y,\sigma = -1,\tau=\pm 1} c_{x,k_y,\sigma=1,\tau=\mp 1}.
\end{equation}
We have restricted the sum over $k_y$ to a single pair of modes, symmetric across $\Gamma$ to preserve $\mathcal{T}'$. At large $x$, $C_K(x) \sim e^{-x/\xi_K}$, where $e^{-1/\xi_K}$ is the largest eigenvalue of the MPS transfer matrix in the relevant charge sector $\Delta Q_K$. Explicitly, $\Delta Q_K = (\Delta q_\mathrm{electric} = 0, \Delta q_\mathrm{valley} = 2, \Delta k_y = 0)$. One way long-range order manifests is a divergence $\xi_K(\chi) \propto \chi$, which we show in Fig.\ref{fig:DMRG}. We caution that the divergence $\xi_K(\chi)$ alone is not enough to uniquely identify the KIVC phase; the TIVC order parameter $C_T(x)$ defined analogously to Eq. \eqref{eq:DMRG_KIVC_order_param} with $O_T^\pm= [\sigma^x \tau^\pm](x,k_y)$, is also governed by the $\Delta Q_K$ sector of the transfer matrix, so a divergence in $\xi_K(\chi)$ could also be a sign of a TIVC phase. If this phase were TIVC, however, then $C_T(x)$ would exhibit long-range order, leading to a large value of $\Sigma_T = \sum_x C_T(x)$, and moreover that value would be governed by the correlation length $\xi_K$. Fig. \ref{fig:KIVC_vs_TIVC} shows this does not occur. Indeed, the TIVC correlator is suppressed by several orders of magnitude relative to the KIVC correlator. We may thus conclude the small strain phase in DMRG is indeed KIVC.

\begin{figure}
    \centering
    \includegraphics[width=0.48\linewidth]{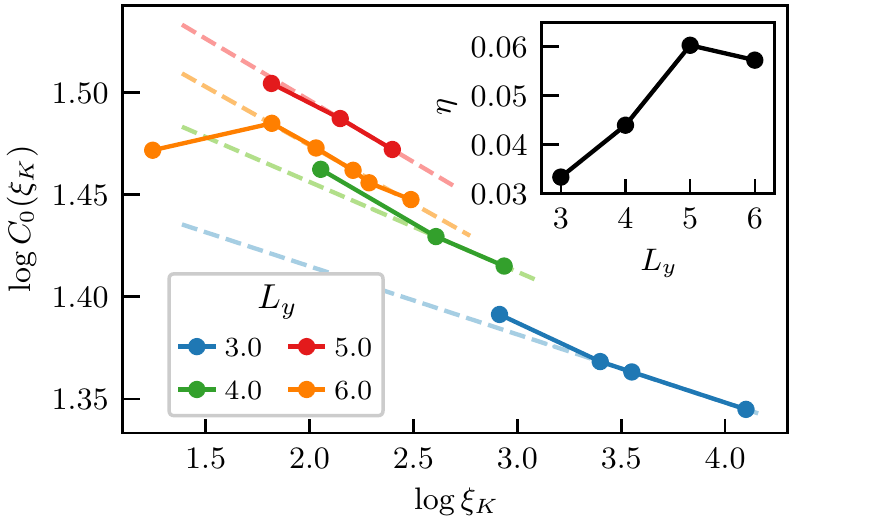}
    \caption{Fitting coefficient $C_0(\xi_K)$ versus $\xi_K$ at a range of cylinder radii $L_y$ at $\nu=0$ and $\epsilon = 0$. One can see that $C_0(\xi_K)$ decays as a power law for sufficiently large $\xi_K$. Inset: fits of $C_0(\xi_K)  = \xi_K^{-\eta(L_y)}$.
    }
    \label{fig:scaling_collapse_intermediate}
\end{figure}

Let us give a few details on how the scaling collapse in Fig. \ref{fig:DMRG} (a) was performed. We expect the KIVC correlator to obey the scaling relation $C_K(x,\xi_K) = \xi_K^{-\eta(L_y)} C_K(x/\xi_K,1)$ for some $\eta(L_y) \xrightarrow{L_y\to \infty} 0$.
Our task is to determine $\eta$ via a fitting procedure to perform the scaling collapse. At a finite bond dimension, all correlators must decay exponentially at sufficiently large $x$ as $e^{-x/\xi_K}$ (perhaps after a regime of algebraic decay). We therefore perform fits $C_K(x\gg 1, \xi_K) = C_0(\xi_K) e^{-x/\xi_K}$. By comparison with the scaling relation, one can see the prefactor to the exponential should scale as $C_0(\xi_K) \approx \xi^{-\eta}$. Fig. \ref{fig:scaling_collapse_intermediate} shows $C_0(\xi_K)$ does indeed decay as a power law for sufficiently large $\xi_K$, and fitting shows $\eta(L_y = 6) \approx 0.057$. A few comments are in order. First, the behavior of $\eta$ is non-monotonic with $L_y$. We attribute this to finite size effects; one requires a cylinder radius of at least $L_y=5$ for KIVC order to be clearly detectable. Second, the magnitude $\eta$ is quite small. Typically, long-range order is visible at finite bond dimension as a range of intermediate $x$ where $C_K(x) \sim x^{-\eta}$. However, as $\eta$ is so small here, this intermediate range is extremely short, so the algebraic decay is not visible in the scaling collapse. Nevertheless, the fact that the data does obey the scaling collapse indicates it must have long-range order.

\subsection{The Semimetal phase in DMRG}

\begin{figure}
    \centering
    \includegraphics[width=0.5\linewidth]{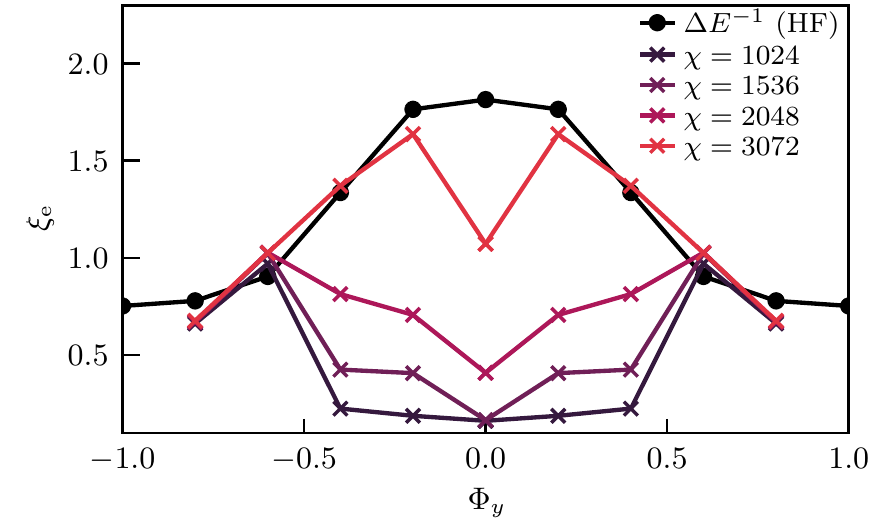}
    \caption{Electronic correlation length $\xi_e$, defined in the text, as the momentum cuts are swept through the mBZ. The black circles are the inverse gap in SCHF at corresponding cuts through the mBZ in arbitary units: $\Delta E^{-1} = \min_{kx} \Delta E(k_x,k_y)$. One can see that the correlation lengths diverge with bond dimension for small $\Phi_y$ where the gap is minimal. Parameters: $\nu =2$, $\epsilon=0.4\%$, $L_y=6$ for DMRG, $L_y = 30$ for SCHF.}
    \label{fig:semimetal_identification}
\end{figure}

We now briefly describe how the semimetal phase found with SCHF can be detected within DMRG. Within DMRG, a phase with zero charge gap has an electronic correlation length which diverges with bond dimension. In this case, we expect a semimetal with two Dirac nodes, so the correlation length will diverge only for particular momenta. As our DMRG uses $L_y=6$ evenly spaced cuts through the moir\'e Brillouin zone at $k_y = 2\pi n/L_y$, we will generically ``miss'' the Dirac nodes, and the correlation length will appear to be finite. However, we can insert (valley-dependent) flux $\Phi_y$, which shifts the cuts to $k_y[n] = 2\pi (n+\tau \Phi_y)/L_y$ where $\tau = \pm 1$ is the valley label. Varying $0 \le \Phi_y \le 1$ will sweep the momentum cuts across the BZ, leading to a divergence in the electronic correlation length when the cut is near to the Dirac node. As SCHF suggests that the Dirac nodes should be quite close to $k_y = 0$, we select the correlator $C_e(x) = \braket{c^\dagger_{x,-k_y[n=0],\sigma=+1,\tau=+1} c^\dagger_{0,k_y[n=0],\sigma=+1,\tau=+1}} \sim e^{-x/\xi_e}$.  Figure \ref{fig:semimetal_identification} shows that the correlation lengths do indeed begin to diverge with $\chi$ precisely where the gap in SCHF is minimal. Together with the fact that $\overline{S}_{vN} \approx 0$ for large $\epsilon$ --- which signals that DMRG finds Slater determinant states --- we may conclude that DMRG detects the same nematic semimetal phase as Hartree-Fock. Indeed, DMRG and SCHF even agree closely on the location of the phase transition from KIVC to semimetal which is around $\epsilon=$ 0.1 \%.

\end{document}